\documentclass[prl,twocolumn,superscriptaddress]{revtex4-1}
\usepackage{latexsym}
\usepackage{amsthm}
\usepackage{amssymb}
\usepackage{amsmath}
\usepackage[all]{xy}
\usepackage{graphicx}
\usepackage{subfigure}
\usepackage{dcolumn}
\usepackage{bm}
\usepackage{bbm}
\usepackage{color}
\usepackage{braket}
\usepackage[usenames,dvipsnames]{xcolor}
\usepackage{amsfonts}
\usepackage{cases}
\usepackage{amssymb}
\usepackage{amsfonts}
\usepackage{bbm}
\usepackage{mathrsfs}
\usepackage{graphics,graphicx,epsfig,bm,amsmath,amsthm,amssymb}
\usepackage{bm}
\usepackage{bbm}
\usepackage{longtable}
\usepackage{multirow}
\usepackage{array}
\usepackage{color}
\usepackage[usenames,dvipsnames]{xcolor}
\usepackage{float}
\usepackage[a4paper,colorlinks=true,
linkcolor=blue,citecolor=blue,
pdfauthor={ },
pdftitle={ },
pdfsubject={ },
pdfkeywords={ }]{hyperref}
\usepackage{leftidx}
\begin{document}

\title{Experimental protection of the coherence of a molecular qubit exceeding a millisecond}

\author{Yingqiu Dai}
\thanks{These authors contributed equally to this work.}
\affiliation{CAS Key Laboratory of Microscale Magnetic Resonance and Department of Modern Physics , University of Science and Technology of China, Hefei 230026, China}
\author{Zhifu Shi}
\thanks{These authors contributed equally to this work.}
\affiliation{CAS Key Laboratory of Microscale Magnetic Resonance and Department of Modern Physics , University of Science and Technology of China, Hefei 230026, China}
\affiliation{Hefei National Laboratory for Physical Sciences at the Microscale, University of Science and Technology of China, Hefei 230026, China}

\author{Yue Fu}
\thanks{These authors contributed equally to this work.}
\affiliation{CAS Key Laboratory of Microscale Magnetic Resonance and Department of Modern Physics , University of Science and Technology of China, Hefei 230026, China}
\affiliation{Hefei National Laboratory for Physical Sciences at the Microscale, University of Science and Technology of China, Hefei 230026, China}

\author{Xi Qin}
\affiliation{CAS Key Laboratory of Microscale Magnetic Resonance and Department of Modern Physics , University of Science and Technology of China, Hefei 230026, China}
\affiliation{Hefei National Laboratory for Physical Sciences at the Microscale, University of Science and Technology of China, Hefei 230026, China}
\affiliation{Synergetic Innovation Center of Quantum Information and Quantum Physics, University of Science and Technology of China, Hefei, 230026, China}

\author{Shiwei Mu}
\affiliation{CAS Key Laboratory of Microscale Magnetic Resonance and Department of Modern Physics , University of Science and Technology of China, Hefei 230026, China}

\author{Yang Wu}
\affiliation{CAS Key Laboratory of Microscale Magnetic Resonance and Department of Modern Physics , University of Science and Technology of China, Hefei 230026, China}

\author{Ji-Hu Su}
\affiliation{CAS Key Laboratory of Microscale Magnetic Resonance and Department of Modern Physics , University of Science and Technology of China, Hefei 230026, China}
\affiliation{Synergetic Innovation Center of Quantum Information and Quantum Physics, University of Science and Technology of China, Hefei, 230026, China}

\author{Lei Qin}
\affiliation{Frontier Institute of Science and Technology (FIST), State Key Laboratory for Mechanical Behavior of Materials, MOE Key Laboratory for Nonequilibrium Synthesis of Condensed Matter and School of Science, Xi¡¯an Jiaotong University, Xi¡¯an 710054, China}

\author{Yuan-Qi Zhai}
\affiliation{Frontier Institute of Science and Technology (FIST), State Key Laboratory for Mechanical Behavior of Materials, MOE Key Laboratory for Nonequilibrium Synthesis of Condensed Matter and School of Science, Xi¡¯an Jiaotong University, Xi¡¯an 710054, China}

\author{Yi-Fei Deng}
\affiliation{Frontier Institute of Science and Technology (FIST), State Key Laboratory for Mechanical Behavior of Materials, MOE Key Laboratory for Nonequilibrium Synthesis of Condensed Matter and School of Science, Xi¡¯an Jiaotong University, Xi¡¯an 710054, China}

\author{Xing Rong}
\email{rong@ustc.edu.cn}
\affiliation{CAS Key Laboratory of Microscale Magnetic Resonance and Department of Modern Physics , University of Science and Technology of China, Hefei 230026, China}
\affiliation{Hefei National Laboratory for Physical Sciences at the Microscale, University of Science and Technology of China, Hefei 230026, China}
\affiliation{Synergetic Innovation Center of Quantum Information and Quantum Physics, University of Science and Technology of China, Hefei, 230026, China}

\author{Jiangfeng Du}
\email{djf@ustc.edu.cn}
\affiliation{CAS Key Laboratory of Microscale Magnetic Resonance and Department of Modern Physics , University of Science and Technology of China, Hefei 230026, China}
\affiliation{Hefei National Laboratory for Physical Sciences at the Microscale, University of Science and Technology of China, Hefei 230026, China}
\affiliation{Synergetic Innovation Center of Quantum Information and Quantum Physics, University of Science and Technology of China, Hefei, 230026, China}


%

\begin{abstract}
 There are several important solid-state systems, such as defects in solids, superconducting circuits and molecular qubits\cite{Nat_QC}, for attractive candidates of quantum computations.
 Molecular qubits, which benefit from the power of chemistry for the tailored and inexpensive synthesis of new systems\cite{ICF,Zheng2016}, face the challenge from decoherence effect.
The decoherence effect is due to the molecular qubits' inevitable interactions to their environment.
Strategies to overcome decoherence effect have been developed, such as designing a nuclear spin free environment\cite{Nat_comm_Bader,JACS} and working at atomic clock transitions\cite{Nat_ACT}.
These chemical approaches, however, have some fundamental limitations.  For example, chemical approach restricts the ligand selection and design to ligands with zero nuclear magnetic dipole moment, such as carbon, oxygen, and sulfur\cite{JACS}.
Herein, a physical approach, named Dynamical decoupling (DD),  is utilized to combat decoherence, while the limitations of the chemical approaches can be avoided.
The phase memory time $T_2$ for a transition metal complex has been prolonged to exceed one millisecond ($1.4~$ms) by employing DD, which is $20$ times longer than that previously reported for such system\cite{Nat_comm_Bader}. 
The single qubit figure of merit $Q_M $ reaches $ 1.4 \times 10^5$. 
Our results show that molecular qubits, with milliseconds $T_2$ , are promising candidates for quantum information processing.
\end{abstract}

\maketitle

Quantum computation provides great speedup over its classical counterpart for certain problems\cite{NielsenQCQI}.
Qubit is the basic element of quantum computation.
There are several candidates for qubits, including superconductor circuits\cite{Nat_Superconduct}, trapped ions\cite{trapped_ions}, defects in solids\cite{Sci_NV} and quantum dots\cite{QD}.
Electron spins in magnetic molecule can be utilized as molecular qubits\cite{Nat_QCinmolecular,ICF}, which recently stimulate great interests.
Compared with other physical systems, molecular qubits have several advantages for building the basic block of quantum computations\cite{ICF}.
All molecules are identical and their structure can be easily tuned by chemical methods.
Molecules can be deposited in regular arrays on surfaces, which is a prerequisite for addressing qubits\cite{ARPC}.
The size of molecular qubit is usually of nanometers, which is suitable for local manipulation and detection\cite{Nature_SS_readout,Science_ThieleS}and is vital for future scalable architecture.
Since the electron spin inevitably interacts with the nearby environment, the quantum coherence of electron spin will lose.
This decoherence effect remains one of main obstacles for molecular qubits towards practical quantum computations.

\begin{figure}[http]
\centering
\includegraphics[width= 1\columnwidth]{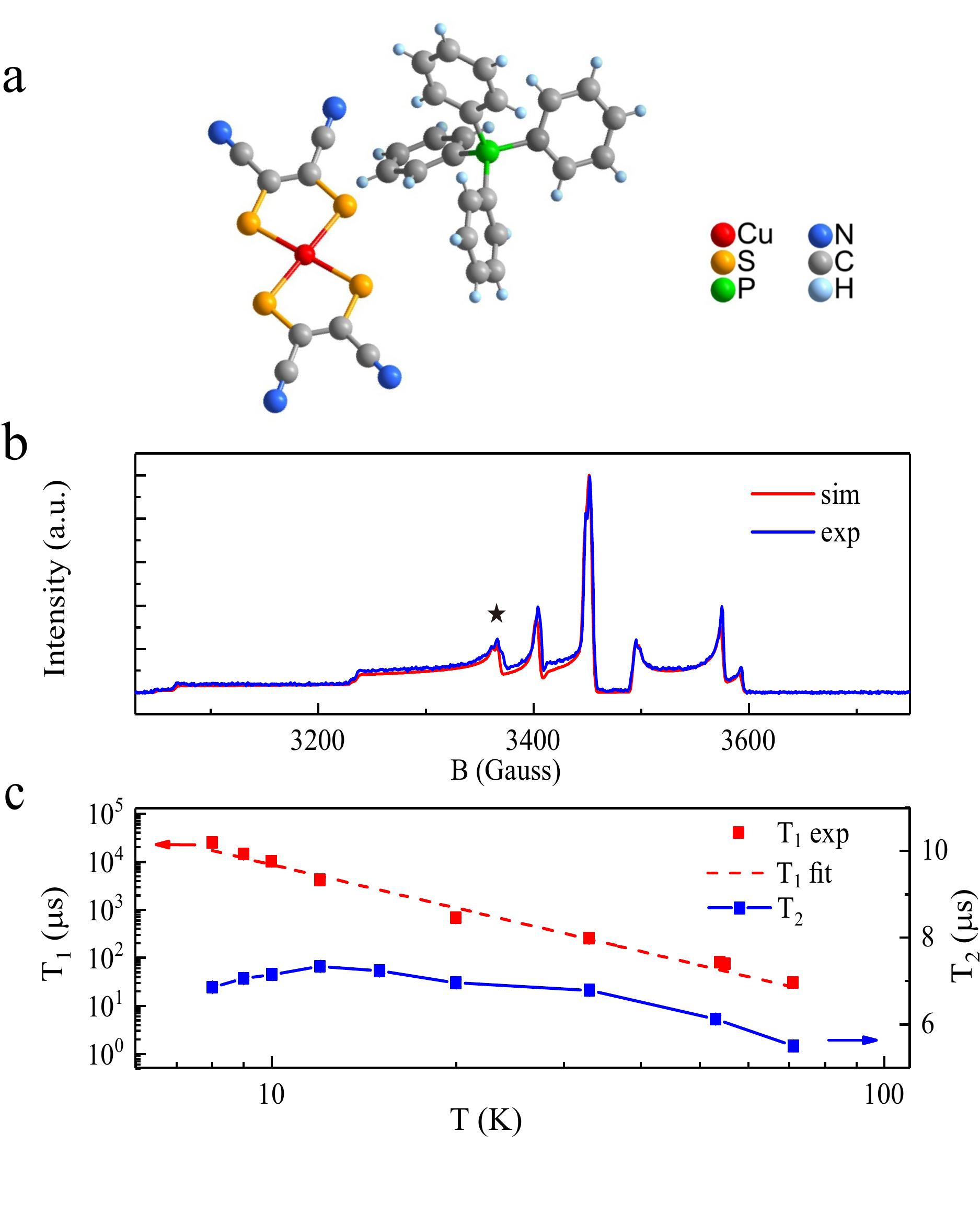}
\caption{ (a) Structure of $\textbf{1Cu}$. Colours: copper-red, sulfur-orange, carbon-grey, nitrogen-blue, phosphorous-green. 
(b) Experimental data of FSED spectrum (blue line) and simulations (red line) with parameters of the Hamiltonian in the main text. (c) Temperature dependence of relaxation times $T_1$ (red rectangles) and $T_2$ (blue rectangles). Red line is fitted according to $1/T_1 \propto T^3$.
}
\label{FIG1_version5}
\end{figure}

Recently, there are several methods developed for molecular qubits to overcome decoherence effect.
One efficient method is to dilute the molecular qubits in a diamagnetic matrix\cite{PRB_Kaminski,RPL_Cr} and enhance the conformation rigidity of the molecule\cite{Nat_comm_Bader}.
Another approach is to design a nuclear spin free environment by synthesizing the chemical complex\cite{Nat_comm_Bader,JACS}.
However, chemical approach restricts the ligand selection and design to ligands with zero nuclear magnetic dipole moment, such as carbon, oxygen, and sulfur.
 One approach is based on the design of molecular structures with crystal field ground states possessing large tunnelling gaps that give rise to atomic clock transitions, where decoherence effect can be suppressed\cite{Nat_ACT}.
These approaches are aiming to prepare the qubits in a nearly noise free or noise insensitive environment, which can be taken as chemical fashion.
The longest reported phase memory time of molecular qubit enhanced by chemical approaches is still below one millisecond\cite{JACS}.
On the other hand, dynamical decoupling technique provides a physical way to fight against the decoherence by modulating qubit's state to average out the noise effect\cite{Nat_ODD}.
With this physical method, there is little limitation in synthesizing the complex.
For example, to replace the hydrogen atoms in the molecular by deuterium\cite{Nat_comm_Bader}  is not necessary for DD.
Another promising advantage of DD is that it can be implemented with quantum gates to execute dephasing suppressed quantum control over qubits\cite{NC_WX,PRL_RX}. 
DD is widely used in quantum dots\cite{PRL_dots}, trapped irons\cite{PRA_irons}, NV centers\cite{NC_NV} and other quantum systems.

A type of transition metal complexes, (PPh$_4$)$_2$[Cu(mnt)$_2$] (\textbf{1Cu}, mnt$^{2-} = $ maleonitriledithiolate) doped into the diamagnetic isostructural host (PPh$_4$)$_2$[Ni(mnt)$_2$] (\textbf{1Ni}), is used as molecular qubits.
The compounds \textbf{1Cu} and \textbf{1Ni} were prepared according to a literature procedure\cite{Nat_comm_Bader}, and were characterized by X-ray crystallography[see section \uppercase\expandafter{\romannumeral2} of appendix  for detail].
The sample in our experiment is diluted in the diamagnetic isostructural host (PPh$_4$)$_2$[Ni(mnt)$_2$]($\textbf{1Ni}$) with concentration $0.3\%$.

Our experiments were performed with an ELEXSYS E580 (X-band) Bruker spectrometer and a homebuilit pulsed EPR spectrometer.
Pulsed EPR experiments that measure temperature dependence of coherence and relaxation time ($T_1$ and $T_2$) were performed with a Bruker Elexsys E580.
Temperatures between 7 and 100~K were achieved with an Oxford Instruments ESR 900 continuous flow helium cryostat.
For prolonging the coherence time with DD, a home-built X band pulsed EPR spectrometer was used.
A $500~$Watt X-band solid-state amplifier with operating frequency range from $8~$GHz to $12~$GHz, which has excellent phase droop and long pulse output up to milliseconds, was equipped in our home-built spectrometer.
An arbitrary sequence generator is utilized to provide the ability to perform multiple pulse DD sequences, which is up to 2048 DD pulses in one sequence in our experiment.
A home-built X-band pulsed-EPR microwave bridge, a commercial Lakeshore's magnet, a Bruker's pulsed ENDOR resonator EN 4118X-MD4 and a continuous flow-type helium X-band EPR cryostat CF935 from Oxford Instruments Ltd were installed in our spectrometer (see section \uppercase\expandafter{\romannumeral1} of appendix  for detail) .


The structure of $\textbf{1Cu}$ is shown in Figure.~\ref{FIG1_version5}a.
There is an electron spin which couples with the nearby copper nuclear spin. The corresponding Hamiltonian can be written as
$H = \textbf{S} \cdot \textbf{A} \cdot \textbf{I} + \beta_e \textbf{B}_0 \cdot \textbf{g}_e \cdot \textbf{S} - \beta_n g_n \textbf{B}_0 \cdot \textbf{I} $
, where $\textbf{S}$ ($\textbf{I}$) stands for the electron (nuclear) spin vector operator, $\beta_e$ ($\beta_n$) stands for the Bohr magneton (nuclear magneton),  $\textbf{g}_e$ is g tensor of the electron spin, $g_n$ is nuclear g factor, $\textbf{B}_0$ is the external magnetic field and $\textbf{A}$ stands for the hyperfine coupling between the electron and nuclear spins.
Figure.~\ref{FIG1_version5}b shows electron spin resonance spectrum of  $\textbf{1Cu}$ by the field sweep electron spin echo detection (FSED) method at temperature $77~$K.
Blue and red lines are experimental and simulation data, respectively.
This spectral shape is due to the anisotropic hyperfine coupling of the electron spin to the $I = 3/2$ copper nuclear spin.
Spectral fits yielded $g_\parallel = 2.0898$, $g_\perp = 2.0215$, $A_\parallel = 495.4~$MHz and $A_\perp = 118~$MHz, which are comparable to the reported data\cite{Nat_comm_Bader}.
The temperature dependencies of relaxation times $T_1$ (spin-lattice relaxation time) and $T_2$ (phase memory time) have been shown in Figure.~\ref{FIG1_version5}c.
The measurements were performed at magnetic field position ($B_0 = 3357~$ Gauss) labeled by the asterisk in Figure. \ref{FIG1_version5}b, which corresponds to the transition between $\ket{m_s,m_I} = \ket {-1/2, 3/2}$  to $\ket {1/2,3/2}$.

\begin{figure}[!htb]

\centering
\includegraphics[width= 1\columnwidth]{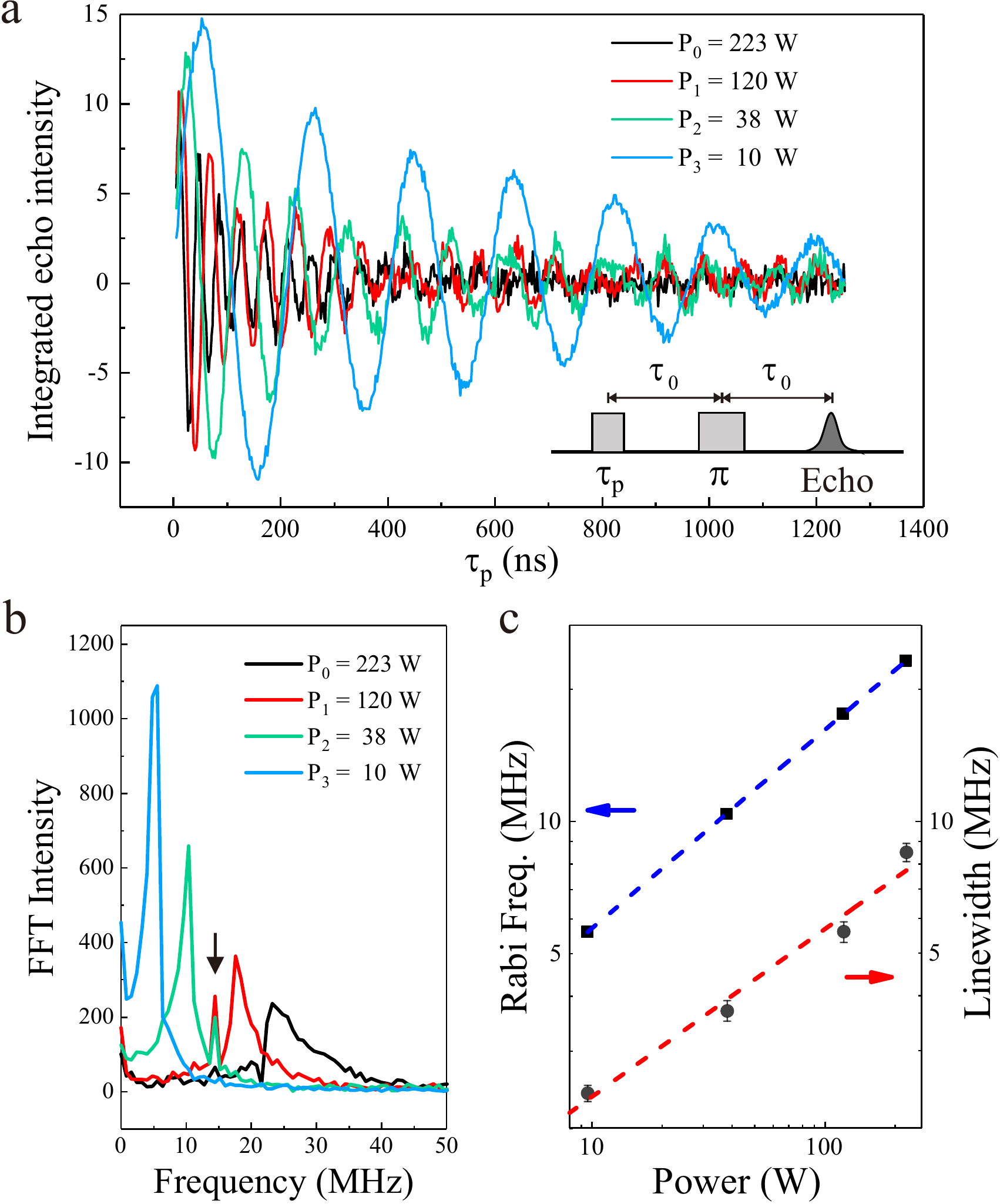}
\caption{ (a) Rabi oscillations for $\textbf{1Cu}$ with different microwave powers recorded at 8~K and $3357~$Gauss. Inset shows the pulse sequence for measuring Rabi oscillations. (b) Fast Fourier transform of Rabi oscillations with different microwave powers. The black arrow marks peaks corresponding to the Larmor frequency of $^{1}H$. (c) Rabi frequency (black rectangles) is proportional to the square root of the power. The linewidth of peaks (black circles with error bars) in (b) is also proportional to the square root of the power.
 }
  \label{Fig2_rabi_version2}
\end{figure}

Spin-lattice relaxation times $T_1$ at different temperatures were measured by inversion recovery sequence $\pi - \tau - \pi/2 - \tau_0 - \pi -\tau_0-$ echo, where $\tau$ stands for the waiting time and $\tau_0 = 400~$ns.
The temperature ranges from $8~$K to $71~$K.
$T_1$ is measured to be $30.4~\mu$s at $71~$K.
With the decreasing temperature, $T_1$ increases dramatically and reaches $25~$ms at $8~$K.
The relaxation rate $1/T_1$ is found to be proportional to $T^3$ (red line in Figure.~\ref{FIG1_version5}c) which displays strong temperature dependence, where $T$ stands for the temperature.
This $T^3$ dependence is attributed to different vibrational frequencies between the local electron spin and the lattices.
The value of $T_2$ also increases with the decreasing temperature and saturates at temperature about $15~$K.
The value of $T_2$ at temperature $8~$K is $6.8~\mu$s, which is longer than $T_2$ at $71~$K with value of $5.5~\mu$s.
With higher temperatures, $T_2$ is bounded by $T_1$, and phonon-induced decoherence is mainly dominated from a two-phonon (Raman) process\cite{PRL_raman}.
For $T = 8~$K, spin relaxation time $T_1$ is enhanced by three orders of magnitude but $T_2 $ is not enhanced a lot.
This indicates that the effect due the surrounding spin-bath\cite{Nat_ODD} becomes the dominant decherence mechanism for this molecular qubit at lower temperatures.

\begin{figure*}[http] 
\centering
\includegraphics[width= 1.8\columnwidth]{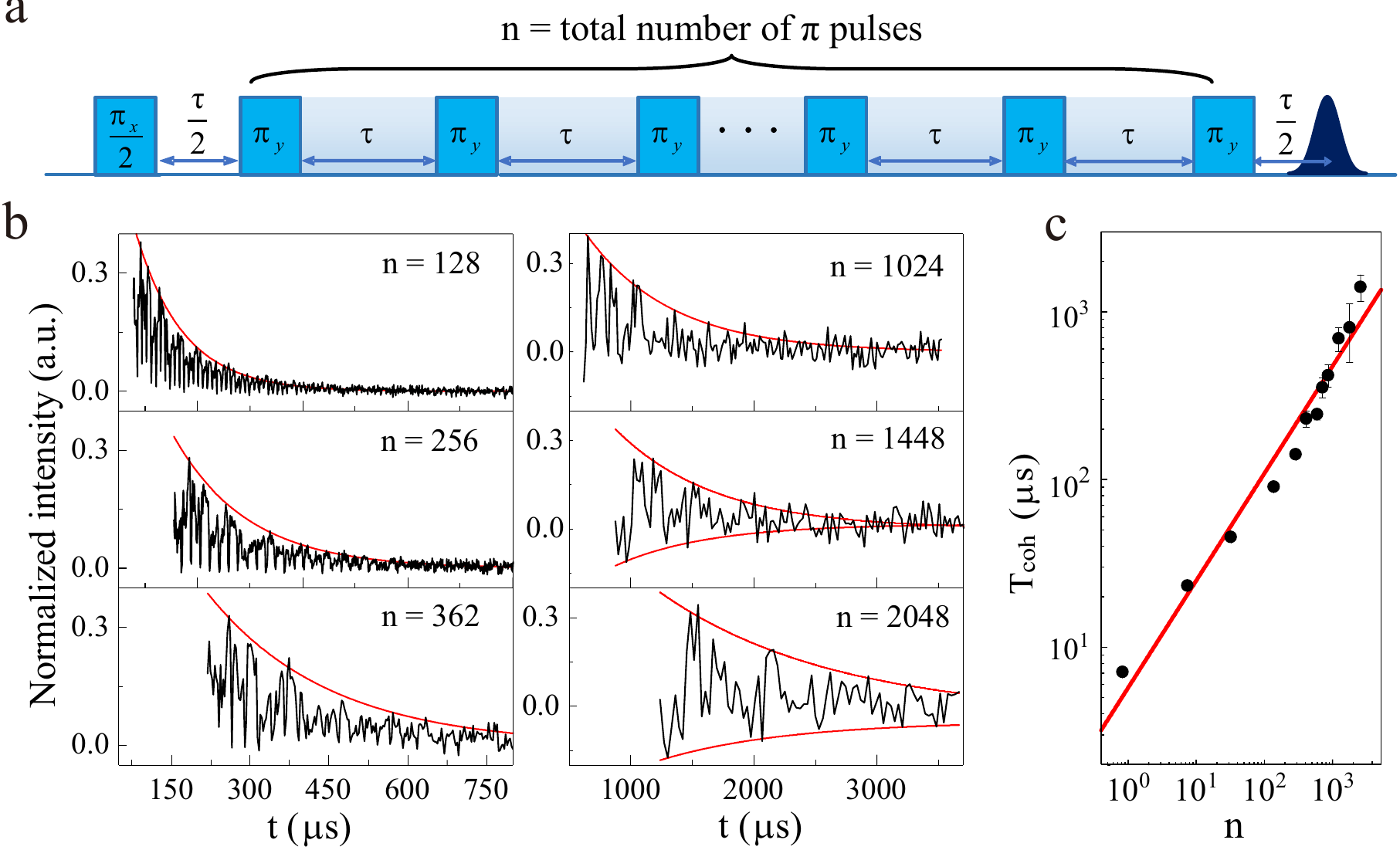}
\caption{(a) Pulse sequence for CPMG-n with applying $n$ times of $\pi$ pulse. $\tau$ stands for the time duration between pulses. (b) Echo intensity decay under different CPMG-n pulse sequences with total evolution time, $t = n\tau$. The coherence times with different CPMG sequences are obtained by fitting the envelopes of echo intensity decay (black lines) to $\exp[-(t / T_{coh})^\beta]$, where $\beta$ is the stretch factor. (c) Coherence times (black circles with error bars) with different number of $\pi$ pulses with scaling $T_{coh}(n) \propto n^{0.67}$ (red line).
}
\label{FIG3}
\end{figure*}

Fig. \ref{Fig2_rabi_version2} shows the experimental Rabi oscillations of this molecular qubit at $8~$K.
Rabi oscillation is to coherently drive the quantum states of electron spin between the two Zeeman splitting energy levels.
It is the basic unit of quantum gates for quantum computations based on spin-systems.
The pulse sequence for Rabi oscillation is $\tau_p - \tau_0 - \pi - \tau_0 - echo$.
It is shown in the inset of Figure. \ref{Fig2_rabi_version2}a, where $\tau_p$ stands for the time duration of the first microwave pulse and the waiting time $\tau_0 = 400~$ns.
The integrated echo intensities dependencies of $\tau_p$ are recorded with different values of input microwave power.
The period of Rabi oscillation becomes shorter with increased power of microwave pulse.
Figure.~\ref{Fig2_rabi_version2}b shows the frequencies of the oscillations obtained by Fourier transforming the data in Figure. \ref{Fig2_rabi_version2}a.
As the power is enhanced, the frequency of Rabi oscillation increases too.
There is an additional frequency observed when the Rabi frequency is closed to the Larmor frequency of protons, which is about $14.3~$MHz at $3357~$Gauss.
This additional frequency is labelled by an arrow in Figure. \ref{Fig2_rabi_version2}b.
This frequency is due to the hyperfine coupling between the electron spin and the nearby protons\cite{PCCP}. 
Figure. \ref{Fig2_rabi_version2}c shows the Rabi frequency and linewidth of peaks dependencies of the microwave power, respectively.
The Rabi frequency is proportional to the square root of the microwave power\cite{APL_rabi}.
The linewidth of peaks with different Rabi frequencies in Figure. \ref{Fig2_rabi_version2}b becomes broader when the power of microwave pulses is enhanced.
This indicates that the decay rate of Rabi oscillation becomes faster with higher driving power.
This can be explained by the static fluctuation of the microwave power\cite{NC_RX}.  
This type of noise effect can be suppressed by composite pulses or quantum optimal control method\cite{NC_RX}.  
Thus high-fidelity quantum gates for molecular qubit can be expected.

Figure.~\ref{FIG3} shows that the phase memory time $T_2$ of this molecular qubit is prolonged by multi-pulse DD sequences at temperature $8~$K.
The DD sequence is an n-pulse Carr-Purcell-Meiboom-Gill (CMPG-n) sequence which is shown in Figure.~\ref{FIG3}a, where $n$ stands for number of refocusing $\pi$ pulses.
This CPMG sequence is $(\pi/ 2)_x - \{\tau/2 - (\pi)_y - \tau/2\}_n - echo$.
The first $(\pi/2)_x$ pulse, which stands for a $\pi/2$ pulse along the $x$ axis in the Bloch sphere, is to generate the quantum coherence.
The following $(\pi)_y$ pulses are to invert the state of the electron spin along the $y$ axis in the Bloch sphere periodically, so that the quasi-static external noise (for example, static external magnetic field fluctuations) can be canceled.
When the number of $\pi$ pulses, $n$, is fixed, the durations between the pulses $\tau$ are varied and then the intensity of echo is recorded.
CPMG sequence is robust against pulse errors, which may come from the fluctuation of the microwave field.
This robustness allows us to apply up to thousands of $\pi$ pluses to protect the quantum coherence.
The other transverse spin component is sensitive to pulse errors under CPMGs and has a much shorter decay time.
This can be improved by other DD sequences (such as XY-8 sequence).
The longitudinal spin component is unaffected by these pulse errors and  decays with a timescale given by $T_1$, which is about $25~$ms for our sample at $8~$K.

Figure.~\ref{FIG3}b shows the experimental result of the preserved electron spin coherence with CPMG sequences.
Number of refocusing $\pi$ pulses ranges from $1$ to $2048$.
The spin coherence can be significantly protected by dynamical decoupling technology.
The spin coherence exhibits a modulated decay behavior as time increases.
The modulation is due to the interaction of electron spin with nearly nuclear spins.
The coherence times with CPMG-n, $T_{coh} (n)$, are obtained by fitting the envelopes of echo intensity (red fitted lines in Figure.~\ref{FIG3}b).
The envelope of echo intensity $M$ follows an exponential decay, $M= \exp[-(t/T_{coh})^\beta]$, where $\beta$ is a parameter determined by the decoherence mechanism of coherence decay.
Figure.~\ref{FIG3}c shows coherence time $T_{coh}$ as a function of DD pulse number $n$.
The coherence time $T_2$ scales as $T_{coh}= T_2 \times n^\alpha$, where $\alpha=0.67 \pm 0.04$ is obtained by fitting the data in Figure. \ref{FIG3}c with $T_2$ as free parameter.
This scaling is similar with the decoherence behavior due to an electron spin bath, which is discussed in Ref.~\onlinecite{NC_NV}.
Because of the broad FSED spectrum (about $550$ Gauss as shown in Figure.~\ref{FIG1_version5}b) and the limited pulse excitation bandwidth (about $100~$MHz with a $10~$ns $\pi/2$ pulse), most of electron spins ($\sim 94\%$) are not excited by microwave pulses.
The off-resonance electron spins behave as an electron spin bath.
The decoherence effect due to this electron spin bath can also be suppressed by dynamical decoupling\cite{NC_NV}.
With 2048 pulses, the phase memory time reaches $1.4 \pm 0.2~$ms, which is the longest one for molecular qubit reported in literatures to the best of our knowledge.
The improvement of the coherence time is not saturated and longer coherence time is expected if more refocusing $\pi$ pulses are applied.
In practice, the performance of DD is limited by the imperfection of the pulses, the minimum time delay between refocusing $\pi$ pulses and the longitudinal relaxation time. There is also a limitation of numbers of pulses due to the solid-state amplifier in our spectrometer.

\begin{table*}\centering  
\newcommand{\tabincell}[2]{\begin{tabular}{@{}#1@{}}#2\end{tabular}}
\caption{\textbf{Comparison between different systems used as qubits.}}
\textrm{\\}
\begin{tabular}{cccc}  

\hline\hline
Systems\qquad \qquad   &  \tabincell{c}{Typical singe-qubit gate \\ operating time }\qquad\qquad   &  Coherence time $T_2$\qquad\qquad  &  $Q_M$\qquad\\ \hline  
Superconducting qubit\cite{NC_SQ}\qquad\qquad   &$1\ $ns \qquad\qquad & $\sim 85\ \mu $s \qquad\qquad     & $8.5\times10^4$\qquad\\
Phosphorous doped in Silicon\cite{NN_SiP}\qquad  \qquad     &$1\ \mu $s\qquad\qquad         & $0.56\ $s \qquad\qquad          & $5\times10^5$\qquad \\
N-V center in diamond\cite{NC_NVtable}\qquad\qquad   &$10\ $ns\ \qquad\qquad          & $0.6\ $s \qquad\qquad           & $6\times10^7$\qquad \\
Trapped ion \cite{PRL_ion}\qquad \qquad & $10\ \mu$s\qquad\qquad  & $50\ $s \qquad\qquad          & $5\times10^6$\qquad \\
Molecular qubit (this work)\qquad \qquad  & $10\ $ns \qquad\qquad & $1.4\ $ms \qquad\qquad        & $1.4\times 10^5$\qquad \\ \hline\hline
\end{tabular}
\label{Table1}
\end{table*}

The dynamics of the coherence under different CPMG sequences reflect fruitful information about the environment.
The modulation in coherence decay comes from the hyperfine couplings between the molecular qubit and the nearby nuclear spins.
As shown in Figure. \ref{FIG3}b, the depth of the coherence modulation increases upon increasing the number of pulses in CPMG sequence.
A model to explain this coherence phenomenon is that the behavior of incoherent nuclear spins around center electron spins could be taken as randomly generated a.c. magnetic fields\cite{NN_dips}.
When dynamical decoupling is applied to the magnetic molecule systems, the a.c. magnetic field will lead to the coherence dips of electron spins at time $t_{dip} = (2k-1)n/2f_l$ where $k$ is dip order and $f_l$ is Larmor frequency of nearby nuclear spins.
According to the result in Figure.~\ref{FIG3}b, dips are induced by $^1H$ nuclear spins.
If more than $1000$ dynamical decoupling pulses are applied on the electron spin, negative spin coherence is observed as shown in Figure.~\ref{FIG3}b.
This can not be explained by the above semiclassical mechanism, which takes the effect of bath spins as classical a.c. magnetic fields.
Because in this picture, the spin coherence is always positive.
This phenomenon can be explained by a quantum mechanism, which treats the environmental nuclear spins as another quantum system.
If the coherence time is prolonged by DD to about several hundreds of microseconds, the effect of the evolution of the bath nuclear spins can not be ignored.
Spin coherence can be expressed by $L(t) = \braket{J_0(t)J_1(t)}$ where $\ket{J_0(t)}$ and $\ket{J_1(t)}$ are nuclear spin states corresponding to electron spin states $\ket{m_S = 1/2}$ and $\ket{m_S = -1/2}$ at time $t$, respectively.
In this quantum decoherence picture, the spin coherence is bounded between $-1$ and $1$ \cite{PRL_Reinhard,NN_dips}.

Table. \ref{Table1} compares several typical physical systems for quantum computations.
Macroscopic systems, such superconducting circuit qubits with micrometer size, provide good tunability, scalability, flexibility and strong coupling to external fields.
However, the coherence time of superconducting circuit is relative short. Recent reported coherence time of SC is about $85~\mu$s\cite{NC_SQ}. The  single qubit figure of merit of SC is about $8.5\times 10^4$ with $1~$ns typical operation time\cite{MRP_SQ}.
On the other hand, microscopic systems, such as spins\cite{NN_SiP,NC_NVtable} and atoms\cite{PRL_ion} with atomic scale, can be utilized as qubits with relative long coherence time (approaching to seconds).
These microscopic systems, however, have limited scalability.
For example, is it difficult to individually address and manipulate many atom qubits\cite{MRP_SQ}.
Molecular qubits, with nanometer size, can be deposited in regular arrays on surfaces, which is prerequisite for addressing qubits, is very attractive for future scalability.
The molecular qubits can be manipulated and readout at single molecular level\cite{Nature_SS_readout, Science_ThieleS}.
The coherence times of molecular have been limited below one milliseconds for many years since molecular qubits have been proposed to execute quantum computations\cite{Nat_QCinmolecular}.

In our experiment, we show for the first time that the coherence time of a type of molecular qubit can be prolonged exceeding one millisecond by dynamical decoupling technique.
It is desirable to compare our result with a previous experiment\cite{Nat_comm_Bader}, which reported coherence time of the same molecular qubit with $68~\mu$s by isotopic purification method.
Though the molecular qubit in our experiment has not been isotopic purified, the coherence time achieved $1.4~$ms by DD, which is twenty folds longer than that reported in Ref. \onlinecite{Nat_comm_Bader}.
Taking the ratio of $T_2$ and the length of $\pi /2$ , a single qubit figure of merit $Q_m$ arrive at $1.4 \times 10^5$ which is useful to evaluate the performance of qubit.
The value of $Q_m$ larger than $10000$ allows for fault tolerant quantum computing\cite{10000}.
%
%
In future, advanced quantum control\cite{NC_RX} can be applied in molecular qubits to perform high-fidelity quantum gates.
Our work marks a important step towards quantum computation with molecular molecule systems.

This work was supported by the 973 Program (Grant No.\ 2013CB921800 and No.\ 2016YFB0501603), the NNSFC (Grant Nos. 11227901, 31470835, 21233007, 21303175, 21322305, 11374305, 11274299, 21473129 and 21620102002), the ``Strategic Priority Research Program (B)'' of the CAS (Grant Nos. XDB01030400 and 01020000). F.S. and X.R. thank the Youth Innovation
Promotion Association of Chinese Academy of Sciences for their support. Y.Z.Z. thanks the support from Wuhan National High Magnetic Field Center (2015KF06) and National 1000 Plan for Young Scholars.

\section*{Appendix}
\subsection {The hardware of the homebuild pulsed ESR spectrometer}

\begin{figure}[!h]
\centering
\includegraphics[width=0.8\columnwidth]{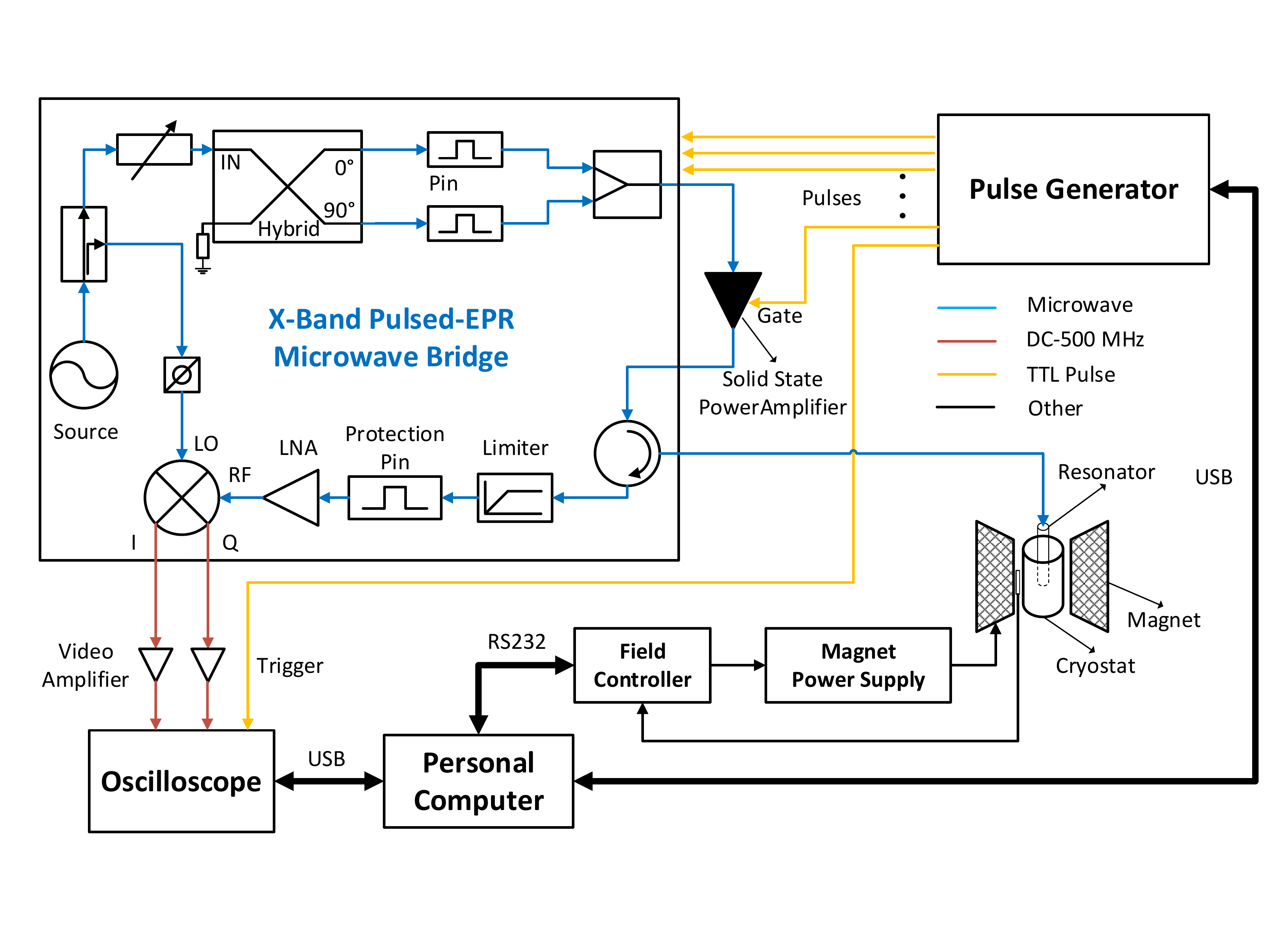}
\caption{Block diagram of the home-built pulsed-EPR spectrometer.}
    \label{FIG S1}
\end{figure}

We designed and constructed the X-band pulsed-EPR spectrometer shown in FIG.~\ref{FIG S1}.
The setup consists of the following parts: microwave bridge, pulse generator, data acquisition device, personal computer, EPR resonator, cryostat system and magnet system.
A pulse sequence generator (ASG, Hefei Quantum Precision Device Co. Ltd., ASG-GT50-C) is used to control microwave pulses and synchronizing the spectrometer.
Keysight's digital oscilloscope MSOS254A with $20~Gsps$ sampling rate and $2.5~GHz$ bandwidth is used as the data acquisition device.
The Pulse Forming Unit has two fixed phases.
A Lakeshore's magnet power supply (Model 665) is used to drive a Lakeshore's electromagnet.
The static field is controlled by Lakeshore's field controller (Model 475 DSP Gaussmeter).
A Bruker's pulsed ENDOR resonator EN 4118X-MD4 is installed in a continuous flow-type helium X-band EPR cryostat CF935 from Oxford Instruments Ltd.
A computer is used to control the equipment through Universal Serial Bus (USB) or RS232 serial communication interface bus.

An X-band 500 Watt solid-state power amplifier (SSPA) is used to amplify the power of microwave pulses in our pulsed-EPR spectrometer.
This SSPA is used instead of the commercial travelling-wave tube amplifier (TWTA) for the following reasons.
The phase of the pulse after the TWTA is distorted, which is known as phase droop.
Phase droop of TWTA leads to poor microwave control pulse fidelity.
The phase droop of SSPA is much better than that of TWTA.
The maximum pulse length with SSPA is about $1~$ms, which is much longer than that of TWAT.
The maximum pulse duration with TWTA is limited to about $15~\mu$s.
Therefore, it is difficult to apply more than $1000$ pulses in one sequence with a TWTA.

The phase droops of both amplifiers are measured.
A rectangular microwave pulse sequence with $0.5\%$ duty cycle is powered by microwave amplifiers.
Then the output signal is demodulated by an IQ demodulator.
The two output signals of the demodulator are collected with a high speed sampling oscilloscope.
Digitized signals of two channels are $I (t) $ and $Q(t)$, where $t$ stands for the time.
The phase is calculated as follows, $\phi (t) = arctan (I(t)/Q(t))$.
For TWTA (ASE 117X) with $1~kW$, $\phi(t)$ changes by $27^{\circ}$ from $t=0$ to $15~\mu s$.
For the solid-state power amplifier, $\phi (t) $ changes by $3^{\circ}$ from $t=0$ to $800~\mu$s.

\begin{figure}[!h]\centering
\includegraphics[width=0.8\columnwidth]{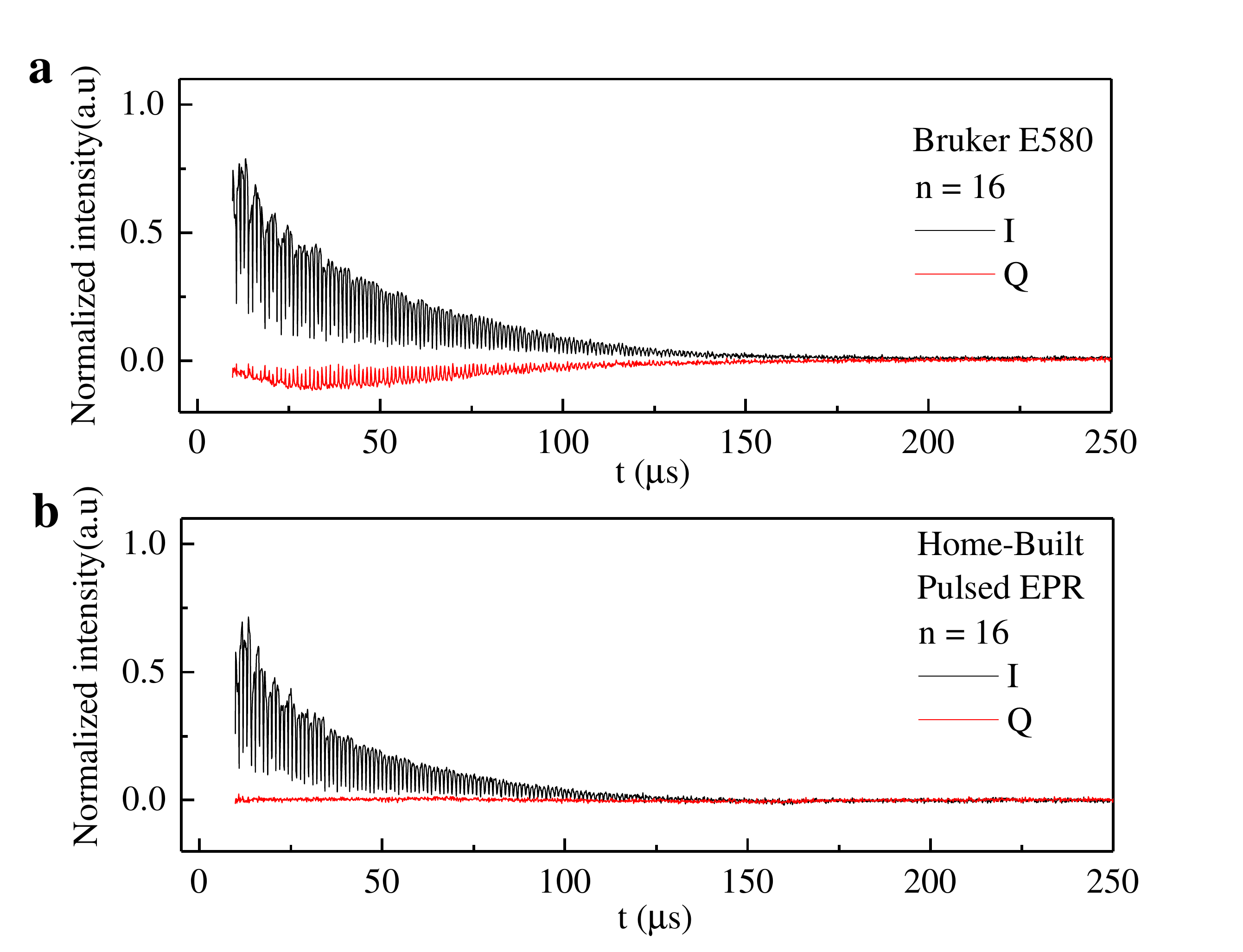}
\caption{ Spin echo decay signals with CPMG-16 detected with TWTA and SSPA, respectively. (a) Spin echo decay is recorded by spectrometer with TWTA.
(b) Spin echo decay is record by spectrometer with SSPA. Black (red) lines stands for the signal acquired by I (Q) channel.}
    \label{FIG S2}
\end{figure}

FIG.~\ref{FIG S2} shows experimental data of $\textbf{Cu1}$ with CPMG-16 obtained by two EPR spectrometers with different power amplifiers.
The outputs from the IQ demodulator correspond to the real and imaginary components of the echo decay signal and are labeled channel $I$ and channel $Q$.
As shown in FIG.~\ref{FIG S2}.a, the decay of spin coherence with TWTA appears in both channels.
This signal with distorted phase is caused by the phase droop of microwave pulses.
The results of CMPG-16 with solid-state power amplifier is shown in FIG.~\ref{FIG S2}.b.
It is clear that our setup with low-phase-droop amplifier provides correct signal.

\subsection*{Sample preparetion}
\begin{figure}[!h]
\centering
\includegraphics[width=0.8\columnwidth]{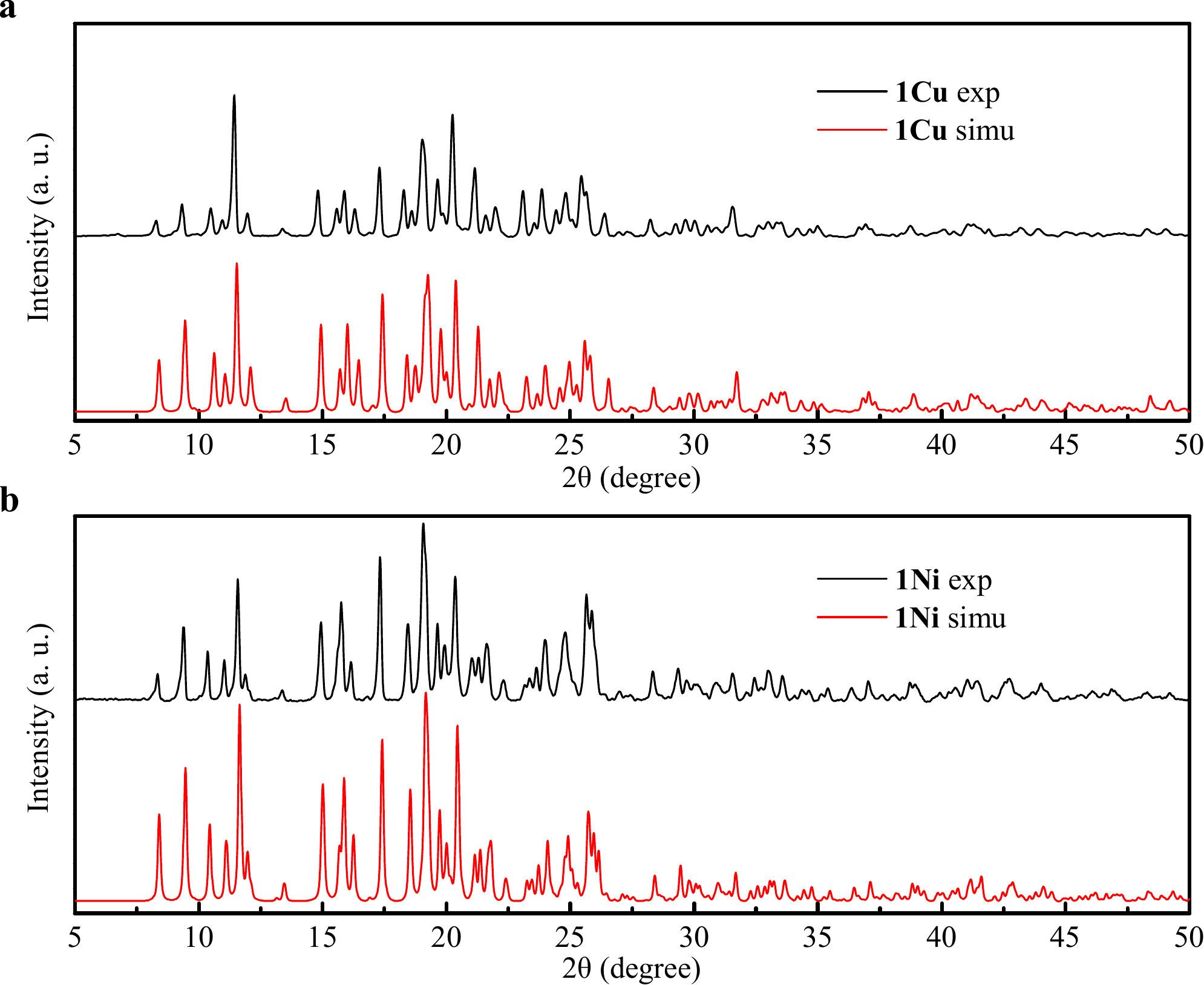}
\caption{XRD spectroscopy. XRD pattern of (a) \textbf{1Cu} and (b) \textbf{1Ni}. Black(red) lines denote the experimental(simulated) results. Experimental data agree with simulations well.
 }
    \label{FIG S3}
\end{figure}

We prepare the compounds \textbf{1Cu} and \textbf{1Ni} according to a literature procedure\cite{Nat_comm_Bader}.
\\Bis-(tetraphenylphosphonium)-bis-(maleonitriledithiolato)cuprate (PPh4)2[Cu(mnt)2] (\textbf{1Cu}): Sodium maleonitriledithiolate (279 mg, 1.5 mmol) was dissolved in 5 ml ethanol and 2 ml demineralized water. Subsequently copper chloride dihydrate (128 mg, 0.75 mmol), dissolved in 5 ml ethanol, and tetraphenylphosphonium bromide (629 mg, 1.50 mmol), dissolved in 15 ml ethanol were added under stirring.
The brown product precipitated immediately and was separated from the solution after 5 min by vacuum filtration. Washing of the product with $3 \times 5$ ml ethanol and drying for 20 h under reduced pressure.
Bis-(tetraphenylphosphonium)-bis-(maleonitriledithiolato)nickelate (PPh4)2[Ni(mnt)2] (\textbf{1Ni}): The same procedure as described for \textbf{1Cu} was executed with nickel chloride hexahydrate (178 mg, 0.75 mmol) instead of copper chloride dihydrate.
Doped powder: 0.3~$\%$ of \textbf{1Cu} in \textbf{1Ni}: Doped powders were obtained by dissolving compounds \textbf{1Cu} and \textbf{1Ni} in the molar ratio 0.3 : 99.97 in a minimum volume of acetone, which was subsequently evaporated under reduced pressure. The resulting powders were dried in vacuo and finely ground.
We characterize the samples by XRD spectroscopy and find that the experimental results are in good agreement with the numerical simulation [See FIG.~\ref{FIG S3}].

\end{document}